\documentclass{iau}
\usepackage{graphicx}
\usepackage{hyperref}
\newcommand{\coh}[2]{\mathsf{{#1}}_{{#2}}}

\title[Bayesian Inference for Radio Observations] 
{Bayesian Inference for Radio Observations - \\Going beyond deconvolution}

\author[M. Lochner, B. Bassett, M. Kunz, I. Natarajan, N. Oozeer, O. Smirnov   \& J. Zwart]   
{Michelle Lochner$^{1,2}$, Bruce Bassett$^{1,2,3}$, Martin Kunz$^{1,4}$, Iniyan Natarajan$^5$, Nadeem Oozeer$^{1,6,7}$, Oleg Smirnov$^{6,8}$ \& Jonathan Zwart$^9$}

\affiliation{$^1$African Institute for Mathematical Sciences, 6 Melrose Road, Muizenberg, 7945, 
South Africa \\email: {\tt dr.michelle.lochner@gmail.com}\\[\affilskip]
$^2$Department of Mathematics and Applied Mathematics, University of Cape Town, Rondebosch, Cape Town, 7700, South Africa\\[\affilskip]
$^3$South African Astronomical Observatory, Observatory Road, Observatory, Cape Town, 7935, South Africa\\[\affilskip]
$^4$D\'epartement de Physique Th\'eorique and Center for Astroparticle Physics, Universit\'e de Gen\`eve, Quai E.\ Ansermet 24, CH-1211 Gen\`eve 4, Switzerland\\[\affilskip]
$^5$Astrophysics, Cosmology and Gravity Centre (ACGC), Department of Astronomy, University of Cape Town, Private Bag X3, Rondebosch 7701, South Africa\\[\affilskip]
$^6$SKA South Africa, 3rd Floor, The Park, Park Road, Pinelands, 7405, South Africa\\[\affilskip]
$^7$Centre for Space Research, North-West University, Potchefstroom 2520, South Africa\\[\affilskip]
$^8$Department of Physics and Electronics, Rhodes University, PO Box 94, Grahamstown, 6140, South Africa\\[\affilskip]
$^9$Department of Physics \& Astronomy, University of the Western Cape, Private Bag X17, Bellville 7535, South Africa}

\pubyear{2014}
\volume{306}  
\pagerange{}
\jname{Statistical Challenges in 21st Century Cosmology}
\editors{A.C. Editor, B.D. Editor \& C.E. Editor, eds.}
\begin{document}

\maketitle

\begin{abstract}
Radio interferometers suffer from the problem of missing information in their data, due to the gaps between the antennas. This results in artifacts, such as bright rings around sources, in the images obtained. Multiple deconvolution algorithms have been proposed to solve this problem and produce cleaner radio images. However, these algorithms are unable to correctly estimate uncertainties in derived scientific parameters or to always include the effects of instrumental errors. We propose an alternative technique called Bayesian Inference for Radio Observations (BIRO) which uses a Bayesian statistical framework to determine the scientific parameters and instrumental errors simultaneously directly from the raw data, without making an image. We use a simple simulation of Westerbork Synthesis Radio Telescope data including pointing errors and beam parameters as instrumental effects, to demonstrate the use of BIRO. 
\end{abstract}

\firstsection 
\section{Introduction}
The problem of extracting scientific parameters from dirty (dominated by artifacts) interferometric radio images has resulted in many deconvolution algorithms being developed. However, none of these solve the problem of incorporating instrumental errors as a source of uncertainty when making measurements from radio data. Deconvolution algorithms can only produce one image, which the scientist must then assume is correct before extracting any science (for example, a catalogue of source fluxes) from it. Algorithms such as CLEAN (\cite{clean}), the most popular algorithm in use, cannot reliably produce any uncertainties (\cite{junklewitz1}), making it impossible to propagate the uncertainties from instrumental errors to the scientific parameters. Further, if these parameters are correlated, as they likely are, even correcting the data for instrumental errors may lead to biased scientific results, as the measurement of the instrumental error may be wrong. These instrumental effects will become more important as more sensitive telescopes such as the Square Kilometre Array\footnote{Square Kilometre Array, \url{http://www.skatelescope.org}} come online. 

\section{BIRO}
We propose in \cite{lochner}, a completely different approach, whereby we model the sky and all known sources of instrumental error simultaneously using the radio interferometry measurement equation (RIME) (\cite{hamaker}). The RIME can be written as (\cite{smirnov}):
\begin{equation}
 \coh{V}{pq} = \boldsymbol{J}_{pn} (\ldots(\boldsymbol{J}_{p2}(\boldsymbol{J}_{p1} \coh{B}{} \boldsymbol{J}_{q1}^H)\boldsymbol{J}_{q2}^H)\ldots)\boldsymbol{J}_{qm}^H,
\end{equation}
where $\coh{V}{pq}$ is the visibility matrix (the radio data), $\boldsymbol{J}_{pi}$ is the $i$'th Jones matrix for antenna $p$ (containing instrumental effects) and $\coh{B}{}$ is the brightness matrix (containing the sky model, and hence all scientific parameters). We use the software package MeqTrees (\cite{noordam}), which implements the RIME, to model our radio field and any known instrumental effects. We can then estimate the parameters of this model, both scientific and instrumental, in a Bayesian context using a sampling method such as MCMC (\cite{metropolis,hastings}). We assume uncorrelated Gaussian noise on the visibilities which leads to a simple Gaussian likelihood for $V_{pq}$.  With this approach, we are able to determine the full posterior for the problem, obtaining not only the best fits for all parameters, but also their uncertainties and correlations.

\section{Applying BIRO}
\begin{figure}[hb]
\centering
 \begin{minipage}{0.45\linewidth}
  \includegraphics[width=0.99\linewidth]{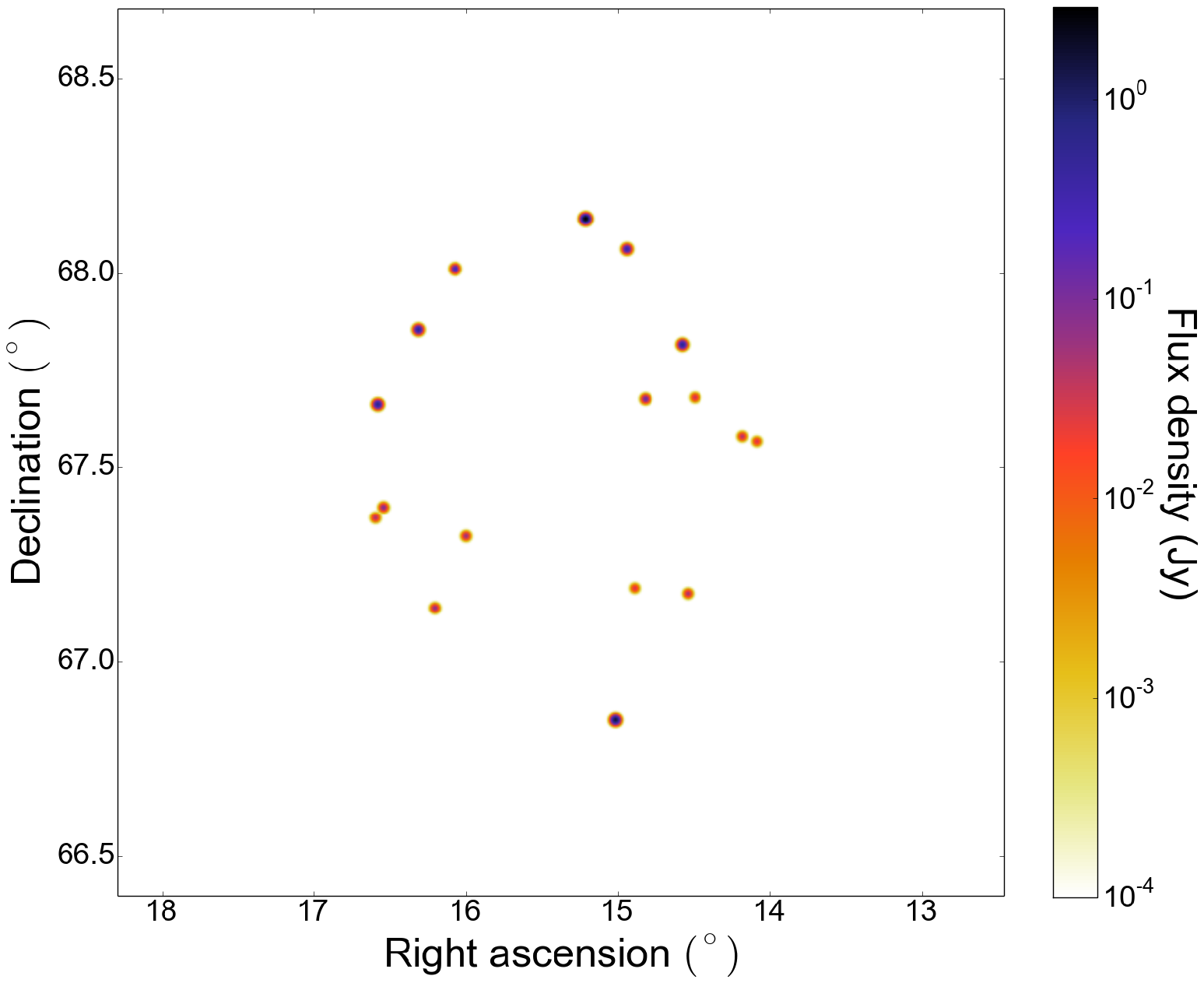}
 \end{minipage}
 \hspace{0.15cm}
 \begin{minipage}{0.45\linewidth}
  \includegraphics[width=0.99\linewidth]{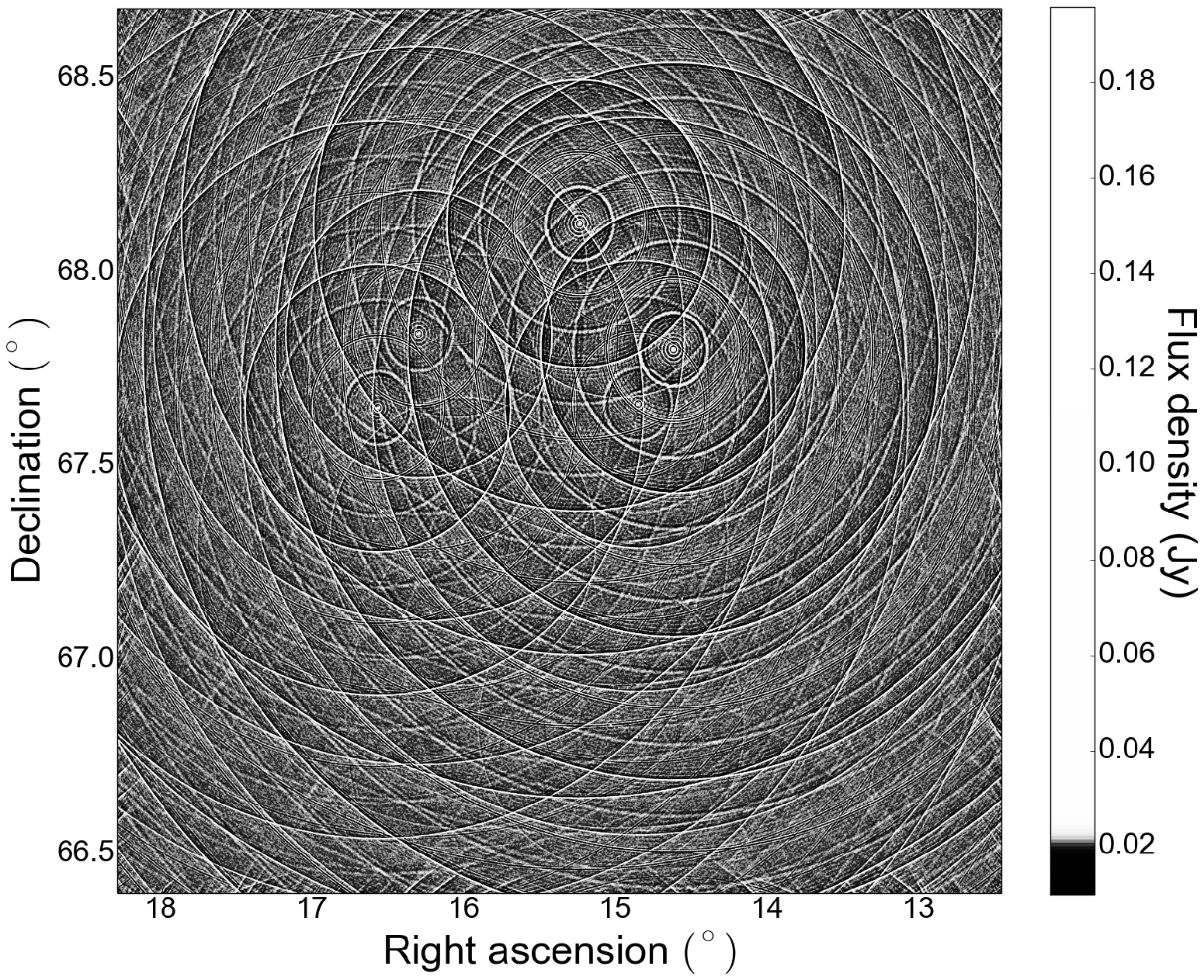}
  
 \end{minipage}
 \caption{Mock dataset. \emph{Left panel:} The simulated sky model. \emph{Right panel:} The ``dirty image'', which is how this field appears when convolved with the telescope beam.}
\label{fig:field}
\end{figure}

Figure \ref{fig:field} shows the mock WSRT\footnote{Westerbork Synthesis Radio Telescope, \url{https://www.astron.nl/radio-observatory/astronomers/wsrt-astronomers}} field which we tested BIRO on. This field, simulated using MeqTrees and based on a real field, consists of 17 point sources. We also applied pointing errors to each antenna as an example of a source of instrumental error, which WSRT (and many other radio telescopes) have had to deal with in the past (\cite{smirnov_pe}). A mispointed antenna will observe a point source through the edge rather than the centre of its beam. Thus, in general, we would expect pointing errors and fluxes to be correlated and we apply our method to determine these correlations, as well as the parameters themselves.

The instrumental parameters for this simulated data consist of the pointing errors, the width of the primary beam and the noise on the visibilities, which we assume to be Gaussian, as is widely considered a good approximation. We allow the pointing errors to vary in time as second ordr polynomial functions, resulting in 84 pointing error parameters, 3 parameters for each direction for each antenna. Our scientific parameters include the flux for each source, and the shape parameters of the extended Gaussian source. 

\begin{figure}[hb]
\centering
 \includegraphics[width=0.5\linewidth]{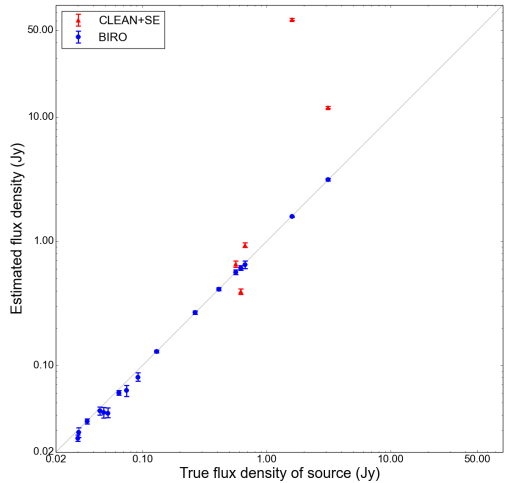}
\caption{Comparison between BIRO and the fluxes obtained from CLEAN, combined with a source extraction algorithm to the data (CLEAN+SE). While the BIRO results are unbiased, CLEAN+SE misestimates the fluxes by up to 44$\sigma$, due to the strong correlations between pointing error and flux.}
 \label{fig:compare}
\end{figure}

\begin{figure}[h!]
 \centering
\includegraphics[width=0.55\textwidth]{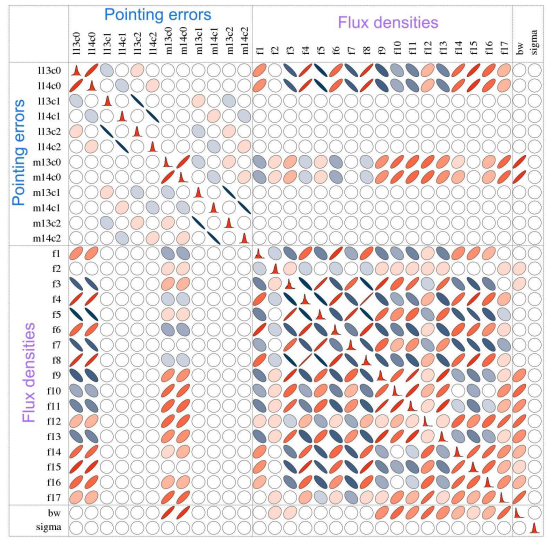}

 \caption{Covariance matrix between a subset of parameters. The parameters are listed on each axis with the correlations between them represented by a coloured ellipse. Highly correlated parameters are red with a thin ellipse angled to the right, whereas anti-correlated parameters have a dark blue ellipse, angled to the left. The diagonal shows the 1D marginalised posterior for each parameter. The correlations in the fluxes arise due to uncertainty in the flux distribution from gaps in the $uv$-plane, while the pointing errors are correlated simply because every pointing error affects every source. Of particular interest is the complex way in which pointing errors correlate with the fluxes, which would be very difficult to determine from first principles.}
 \label{fig:covmat}
\end{figure}

We compared BIRO with the standard, commonly used CLEAN algorithm combined with a source extraction algorithm on the CLEANed image (we call this CLEAN+SE) for this dataset. Figure \ref{fig:compare} illustrates this comparison for the estimated fluxes from BIRO and CLEAN+SE. With no knowledge of the pointing errors, it is no surprise that CLEAN+SE returns biased fluxes and fails to find several sources. The danger is that the error bars, estimated using only the surrounding flux of the point source in this source extraction algorithm, do not and cannot take into account the additional source of uncertainty from the instrumental errors. In contrast, the BIRO estimates are unbiased and the error bars are larger, correctly propagating the instrumental errors, and determining the correlations between parameters (Figure \ref{fig:covmat}).

\section{Conclusions}
We have introduced BIRO, a Bayesian approach to the deconvolution problem of radio interferometry observations. Figure \ref{fig:covmat} highlights the importance of fully propagating the uncertainty on instrumental errors, as they can be highly correlated with the scientific parameters and hence bias scientific results. Due to the fully Bayesian nature of BIRO, it allows for very elegant extensions to this simple scenario. For example, the choice of model for the field can be selected for using the Bayesian evidence (see \cite{lochner}). BIRO can be useful in any scenario where reliable statistics are required for the science extracted and may be essential to fully exploit the sensitivity of the SKA.

\section{Acknowlegdements}
ML and JZ are grateful to the South Africa National Research Foundation Square Kilometre Array Project for financial support. ML acknowledges support from the University of Cape Town and resources from the African Institute for Mathematical Sciences. OS is supported by the South African Research Chairs Initiative of the Department of Science and Technology and National Research Foundation. IN acknowledges the MeerKAT HPC for Radio Astronomy Programme. Part of the computations were performed using facilities provided by the University of Cape Town's ICTS High Performance Computing team: \url{http://hpc.uct.ac.za}.
\end{document}